\documentclass[journal]{IEEEtran}

\usepackage{cite}

\ifCLASSINFOpdf
   \usepackage[pdftex]{graphicx}
   \graphicspath{{../pdf/}{../jpeg/}}
   \DeclareGraphicsExtensions{.pdf,.jpeg,.png}
\else
\fi
\usepackage{amsmath}
\usepackage{mathrsfs}

\usepackage{array}

\usepackage{subfig}
\usepackage{graphicx}
\usepackage{cite}
\newtheorem{theorem}{Theorem}[section]

\newtheorem{lemma}[theorem]{Lemma}
\newtheorem{property}[theorem]{Property}
\newtheorem{definition}[theorem]{Definition}

\usepackage{arydshln}
\usepackage{lipsum}
\usepackage{multicol,blindtext}
\usepackage{mathtools, cuted}
\usepackage{lipsum, color}
\usepackage{algorithm}
\usepackage[noend]{algpseudocode}
\makeatletter
\def\BState{\State\hskip-\ALG@thistlm}

\makeatother

\usepackage{url}

\begin{document}
\title{Data-Driven Wide-Area Control Design of Power System Using the Passivity Shortage Framework}

\author{Ying~Xu,~\IEEEmembership{Member,~IEEE,}
	    Zhihua~Qu,~\IEEEmembership{Fellow,~IEEE,}
    	Roland~Harvey,~\IEEEmembership{Student Member,~IEEE,}
        and Toru Namerikawa,~\IEEEmembership{Member,~IEEE,}
\thanks{Y. Xu, Z. Qu and R. Harvey are with Department of Electrical and Computer Engineering, University of Central Florida, Orlando 32816, USA.
       Emails: {\sl ying.xu@ucf.edu}, {\sl qu@ucf.edu}.
       This work is supported in part by US National Science Foundation under grants ECCS-1308928 and ECCS-1552073, by US Department of Energy's awards DE-EE0006340, DE-EE0007327 and DE-EE0007998, by Leidos' contract P010161530, and by Texas Instruments' grants. T. Namerikawa is with Keio University, Japan.}}

\markboth{IEEE Transactions on Power Systems: completed on \today}%
{Shell \MakeLowercase{\textit{et al.}}: Bare Demo of IEEEtran.cls for IEEE Journals}

\maketitle

\begin{abstract}
A novel wide-area control design is presented to mitigate inter-area power frequency oscillations. A large-scale power system is decomposed into a network of passivity-short subsystems whose nonlinear interconnections have a state-dependent affine form, and by utilizing the passivity shortage framework a two-level design procedure is developed. At the lower level, any generator control can be viewed as one that makes the generator passivity-short and $L_2$ stable, and the stability impact of the lower-level control on the overall system can be characterized in terms of two parameters. While the system is nonlinear, the impact parameters can be optimized by solving a data-driven matrix inequality (DMI), and the high-level wide-area control is then designed by solving another Lyapunov matrix inequality in terms of the design parameters. The proposed methodology makes the design modular, and the resulting control is adaptive with respect to operating conditions of the power system. A test system is used to illustrate the proposed design, including DMI and the wide-area control, and simulation results demonstrate effectiveness in damping out inter-area oscillations.
\end{abstract}

\begin{IEEEkeywords}
wide-area control, data-driven control, matrix inequality, Lyapunov stability, passivity-short systems, power systems
\end{IEEEkeywords}

%
\IEEEpeerreviewmaketitle

\section{Introduction}

\IEEEPARstart{I}{nter-area} oscillations observed in large-scale power systems are typically recognized as low frequency problems on the order of 0.1-–1.0 Hz. As the system expands and energy interchanges between interconnected systems increase, these low-frequency inter-area oscillations often become poorly damped. Recently, this problem has been even more challenging due to the fast development and high penetration of renewable resources. To solve this problem, tremendous effort has been made in the past decades.

In the traditional damping design, each subsystem is treated as an independent control, and each one is capable of acting on its own. For example, using power system stabilizers (PSS) is a typical local control design, forming an additional part of the generation control system. However, it is well-known that local designs may not always be effective to damp out the inter-area modes of oscillations for the following reasons:
\begin{itemize}
\item The design is usually based on the internalization of each individual subsystem under certain operation conditions, thus, the stability may not be ensured under any local design as the operation point of a power system changes.
\item Although an overall centralized control (e.g. AGC) has been used in power systems for years, a systematic design has not been reported yet to assure the overall stability.
\end{itemize}

With the advent of time-synchronized phasor measurement units (PMUs) and fast-speed communication technologies, the concept of wide-area measurement systems (WAMS) has attracted research interest and has become indispensable in addressing such issues as instability detection and control, security assessment and enhancement in modern power systems.

Progress has been reported in wide-area control of power systems, and detailed results on improving inter-area oscillation damping are presented in \cite{Zhang16sta,Shen17Ada,Yousefian17Hyb,ma14app,Soloot15wide,li17design,bi17impact,Zenelis19wide,Liu18osc}. Most of the existing literature on this topic usually focus upon one or a few of the following aspects: power system model recognition, wide-area damping controller design, and parameter tuning methodology and performance validation.

The difficulty of processing the large amount of data captured by WAMS over geographically dispersed locations has been one of the major issues in the application of PMU data. In \cite{vahidnia2015wide}, inter-area dynamics of the overall system are represented by a reduced-order model based on the estimation of aggregated system angles and velocities by using a non-linear Kalman filter. {blue}{A dynamic  eigensystem  realization algorithm is presented in \cite{dobrowolski2018inter} to identify the inter-area oscillation mode, and based on that a linear quadratic Gaussian controller is implemented through PSS. A method using PMU measurements on specific points is presented in \cite{chakrabortty2011} to construct dynamic inter-area system models by aggregating the generators inside each area. This approach establishes feasibility of the wide-area control through aggregating groups of generators.}

Several approaches based on robust control theories and linear matrix inequalities (LMIs) have been applied to wide-area damping control designs \cite{zhang2008design,ni2002power}. For instance in \cite{zhang2008design}, a general procedure is proposed to design a wide-area damping controller by applying an LMI approach to the problem of regional pole placements. Designs based on artificial neural networks has become more popular recently, but those methods require an appropriate set of training data \cite{6030909}. A model free method is presented in \cite{2018arXiv180409827F} by applying online reinforcement learning as the data-driven wide-area oscillation-damping control.

Although the aforementioned results present significant progress, there are unresolved issues that limit the application and performance of wide-area control in actual power systems. The most critical issue to be addressed is an overall stability analysis of interconnected dynamic systems. To this end, a novel systematic control design is proposed in this paper, and it uses the framework of passivity shortage, outlined in \cite{harvey2016cooperative}, applied to power systems. It should be noted that interconnection of passive systems was studied in \cite{arcak2007passivity}, \cite{chopra2006} among others but, in power systems, devices such as synchronous generators are not passive. Hence, passivity-short systems and their properties must be applied. It is worth recalling that passivity-short systems and their properties are investigated in \cite{qu2014modularized}, and generator dynamics are always passivity-short \cite{harvey2017dissipativity}. In this paper, this energy-based approach is applied to design a two-level control by taking advantage of wide-area measurement data.

Using the passivity-shortage framework, stability of the overall system is investigated as the interconnection of {blue}{subsystems or groups of coherent generators, their individual generator controls, and their wide-area control}. By nature, each of the subsystems is passivity-short and $L_2$ stable. With WAMS data, reduced-order load flow equations can be identified, and the impact of passivity-short {blue}{subsystems of coherent generators and their interconnections can be quantified by two parameters} using data-driven matrix inequalities. As such, their impacts can be minimized by the design of individual controls for the subsystems. And, the high-level control can then be synthesized to ensure the overall system stability and hence to effectively damp out potential inter-area oscillations. The proposed data-driven control design allows the controls to adapt themselves to both the power system operating conditions and their transient behaviors, resulting in superior performance.

The remainder of this paper is organized as follows. In section II, {blue}{(aggregated) models of  synchronous and renewable generators are cast into the passivity-short framework}, and a two-level control design problem is formulated. In section III, a four-step design process is presented for both local and wide-area controls, data driven optimization is done, and rigorous analysis of the overall system is performed. In section IV, communication topology, the control design algorithm, and its robustness are discussed. In section V, simulation results are presented to illustrate the effectiveness of the proposed design, followed by concluding remarks in section VI.

\section{Problem Formulation}

Consider the class of interconnected dynamic systems described by the following heterogeneous nonlinear model: for the $i$th subsystem,
\begin{equation}
\begin{split}
\dot x_i &= A_i(x_i) x_i + B_i(x_i) v_i + \sum_{j\in {\cal N}_i} H_{ij}(y_i,y_j) (y_j-y_i),
\\ y_i&=C_i(x_i)x_i,
\end{split}
\label{generalsystem}
\end{equation}
where ${\cal N}=\{1,\cdots,n\}$ is the index set, $i\in {\cal N}$,
${\cal N}_i\subset {\cal N} $ denotes the neighboring set of system $i$, $x_i \in\Re^{n_{g_i}}$ is the state vector of the $i$th subsystem, $y_i\in\Re^{l}$ is its output vector, $v_i\in \Re^{m}$ is its control input, matrices $A_{i}(x_i)$, $B_i(x_i)$, $C_i(x_i)$ are state-dependent and of proper dimensions, and coupling matrices $H_{ij}(y_i,y_j)$ are output-dependent and of proper dimensions.

It is shown in appendix \ref{sec:modeling} that dynamics of a power system of $n$ generators, {blue}{which could be either conventional or renewables or their aggregated models in certain geographical regions}, are described by a set of nonlinear differential-algebraic equations. Those equations can be recast into system equations in the form of \eqref{generalsystem} and with respect to their equilibrium. In the appendix, vectors $x_i$, $y_i$, and $v_i$ as well as matrices $A_{i}$, $B_i$, and $C_i$ are detailed so the proposed control design becomes directly applicable. Specifically, the state/output in \eqref{eq:xi} and the matrices in \eqref{eq:ABCH} should be used for conventional generators, while inverter-based energy sources have the state/output and their corresponding matrices defined by \eqref{eq:xi_der} and \eqref{eq:ABCH_der}, respectively.

For the $i$th system in \eqref{generalsystem}, the proposed wide-area control is of form
\begin{align}
v_i = -K_ix_i + u_i, \label{eq:con}
\end{align}
where the first term $-K_ix_i$ is the self-feedback control\footnote{If needed, a nonlinear self-feedback control of form $-K_i(x_i)x_i$ could be designed.}, and $u_i$ is the WAMS-enabled control as
\begin{align}
u_i = - k_{c_i} \sum_{j\in {\cal N}_{c_i}} S_{ij}^c(t) (y_j-y_i),
\label{eq:wa-con}
\end{align}
in which $k_{c_i}>0$ is a control gain, and $S_{ij}^c(t)$ denotes the possibly time-varying communication matrix for WAMS: given communication neighboring set ${\cal N}_{c_i}(t)$ of the $i$th subsystem,
\begin{equation}
S_{ij}^c(t) =
\begin{cases}
1&\text{if}\;\;j \in {\cal N}_{c_i}\\
0&\text{otherwise}
\end{cases}. \nonumber
\end{equation}
Under control \eqref{eq:con}, system \eqref{generalsystem} can be rewritten as
\begin{equation}
\begin{split}
\dot x_i &= \overline{A}_i(x_i) x_i + B_i(x_i) u_i + \sum_{j\in {\cal N}_i} H_{ij}(y_i,y_j) (y_j-y_i),
\\ y_i&=C_i(x_i)x_i,
\end{split}
\label{generalsystem-i}
\end{equation}
where $\overline{A}_i(x_i)=A_i(x_i)-B_i(x_i)K_i$. Its nominal controlled dynamics (excluding the interconnection) are expressed as
\begin{equation}
\begin{split}
\dot x_i &= \overline{A}_i(x_i) x_i + B_i(x_i) u_i,
\\ y_i&=C_i(x_i)x_i.
\end{split}
\label{nominalsystem}
\end{equation}
Design of control \eqref{eq:con} for all the subsystems involves choices of feedback gain matrices $K_i$, communication neighboring sets ${\cal N}_{c_i}$, and cooperative control gain matrix $K_c=\mbox{diag}\{k_{c_i}\}$. Feedback gain matrices and the cooperative control gain matrix will be synthesized in section \ref{DDC1}, and choices of communication topology will be discussed in section \ref{sec:comm}.

The proposed design employs two novel tools. One is the analytical framework of passivity-short dynamic systems, as summarized in the following definition. The concept of passivity-short systems is used because most physical systems, including the swing equation in power systems, are not passive.

\noindent \begin{definition} Nominal subsystem \eqref{nominalsystem}, or its input-output pair $(u_i,y_i)$, is said to be {\it passivity-short} with respect to storage function $V_i$ if inequality
\begin{equation}
\dot V_i \leq u_i^Ty_i + \frac{\epsilon_{ii}}{2}\|u_i\|^2 -\frac{\rho_i}{2}\|y_i\|^2,\label{PS}
\end{equation}
holds for some $\epsilon_{ii}>0$. The subsystem is said to be {\it $L_2$ stable} if inequality \eqref{PS} holds for some $\rho_i>0$ and a positive definite $V_i$. The subsystem is said to be {\it passive} if inequality \eqref{PS} holds for some $\epsilon_{ii}=0$ (and $\rho_i=0$).
\label{definition}
\end{definition}

The second tool employed is an efficient computational algorithm involving {\it data-driven matrix inequality} (DMI). To illustrate the idea, consider nominal subsystem \eqref{nominalsystem} and suppose that the storage function $V_i$ in Definition \ref{definition} is chosen to be quadratic as
\begin{equation}
V_i = \frac{1}{2} x_i^TP_ix_i,
\label{fcnv}
\end{equation}
where $P_i$ is a positive definite matrix. Then, we know from Lyapunov's direct method that the nominal subsystem is passivity-short provided that
\[
\left[ \begin{array}{c:c}
\overline{A}_i^T P_i+P_i \overline{A}_i + \rho_i C_i^TC_i & P_i B_i - C_i^T \\ \hdashline
B_i^TP_i - C_i & - \epsilon_{ii}I \end{array} \right]<0. \]
By Schur complement lemma \cite{vanantwerp2000tutorial}, the above matrix inequality is equivalent to
\begin{equation}
M_i(x_i) \stackrel{\triangle}{=}\overline A_i^T P_i + P_i \overline A_i + \rho_i C_i^T C_i + \frac{1}{\epsilon_{ii}}\|P_i B_i-C_i^T\|^2
<0. \label{DMI}
\end{equation}
Matrix inequality \eqref{DMI} is state-dependent but can efficiently be solved real-time for $K_i$ (within matrix $\overline{A}_i$), $\epsilon_{ii}$ and $\rho_i$. This design based on data-driven matrix inequality \eqref{DMI} not only applies to individual subsystems but also to the interconnected system as a whole, and it also makes it possible to modularly synthesize a multi-level control for the resulting system, as shown by the four-step design process outlined in the subsequent section.

\section{Modular Control Design}\label{DDC1}

In this section, a modular control design is presented for the interconnected system consisting of \eqref{eq:con}, \eqref{eq:wa-con}, and \eqref{generalsystem-i}. The first step is to determine $H_{ij}$ using real-time measurement data. The second step is to design individual feedback control gain $K_i$ so that individual nominal system \eqref{nominalsystem} passivity-short. The third step is to quantify the impact of interconnections among the subsystems on stability of the overall system. The fourth step is to make the overall system synchronize by appropriately synthesizing the wide-area control \eqref{eq:wa-con}. When applied to power systems, steps two to four quantitatively prescribe the impacts of generator controls, load flow equations, and wide-area control on the overall power system stability and performance, respectively. Design steps two and three are expressed and solved as a real-time optimization problem so the overall system performance is ensured and enhanced.

\subsection{Data-Driven Calculation of $H_{ij}$ \label{sec:H}}

{blue}{Assuming that the (aggregated) power system has $N_b$ buses, we know that power flow equations are} described by
\begin{eqnarray*}
&& P_i - P_{L_i} = \sum_{j=1}^{N_b} V_{i}V_{j}(g_{ij} \cos{\theta_{ij}}+b_{ij}\sin{\theta_{ij}}), \\
&& Q_i- Q_{L_i} = \sum_{j=1}^{N_b} V_{i}V_{j}(g_{ij} \sin{\theta_{ij}}-b_{ij}\cos{\theta_{ij}}),
\end{eqnarray*}\noindent
where $V_i$ is the nodal voltage, $\theta_i$ is the phase angle, $P_i$ is the power injection, $P_{L_i}$ is the load, all at the $i$th bus; $\{g_{ij}, b_{ij}\}$ are the real- and imaginary-part of the $(i,j)$th element in the power network admittance matrix, and $\theta_{ij}=\theta_i-\theta_j$. Due to the expansive nature of power transmission/distribution networks, it is impossible to monitor all the bus voltages, load variations, and topology changes (i.e., parameter variations of $g_{ij}$ and $b_{ij}$) within the overall system.

With the development of WAMS, phase angles ($\delta_i$) and power injections ($P_{g_i}, Q_{g_i}$) {blue}{at all geographical regions (i.e. major groups of power generation units)} are monitored real-time. Accordingly, a reduced-order set of power equations can equivalently be established at the power generation level as follows:
\begin{subequations}
	\begin{equation}
	P_{g_i} = \sum_{j \in {\cal N}} E_{i}E_{j}(G_{ij} \cos{\delta_{ij}}+B_{ij}\sin{\delta_{ij}}),\label{eq:gi-flow}\\
	\end{equation}
	\begin{equation}
	Q_{g_i} = \sum_{j \in {\cal N}} E_{i}E_{j}(G_{ij} \sin{\delta_{ij}}-B_{ij}\cos{\delta_{ij}}), \label{eq:gi-pf}
\end{equation}\label{eq:gi}
\end{subequations}\noindent
\hspace*{-0.14in} where $E_i$ is the inner bus voltage behind transient reactance of the $i$th generator,
$\{P_{g_i}, Q_{g_i}\}$ are the active and reactive power injections by the $i$th generator,
$G_{ij}$ and $B_{ij}$ are the real and imaginary parts of the $(i, j)$th entry in the reduced-order network admittance matrix, and $\delta_{ij}=\delta_{i}-\delta_{j}$ is the angle difference between the $i$th and $j$th generators. {blue}{By collecting recent time series measurement of $E_i$, $\delta_{ij}$, $P_{g_i}$ and $Q_{g_i}$, parameters $G_{ij}$ and $B_{ij}$ can be estimated by applying such standard techniques as the least square method \cite{fan2013} to equivalent network equations in \eqref{eq:gi}. Although real-time estimation is possible, system topology may change, and estimation error may not be neglectable. Hence, the proposed control design only requires the ranges of parameters $G_{ij}$ and $B_{ij}$ rather than not their accurate estimates, as shown below and in section IV.C.}

It follows from \eqref{eq:ABCH} or \eqref{eq:ABCH_der} in appendix A that
\[ H_{ij}(y_i,y_j)=\left[ \begin{array}{cc} {\bf 0} & {\bf 0} \\\hdashline  \frac{h_{ij}(y_{i1},y_{j1})}{M_i}& 0  \\ \hdashline  {\bf 0} & {\bf 0} \end{array} \right], \]
where $y_{i1}=\delta_{i}-\delta_{i}^*$, $\delta_i^*$ is the equilibrium of $\delta_i$, and
\begin{equation}
h_{ij}  = E_iE_j\frac{G_{ij}(\cos \delta_{ij}- \cos \delta_{ij}^*) + B_{ij}(\sin \delta_{ij}- \sin \delta_{ij}^*)}{\delta_{ij}-\delta_{ij}^*}, \label{eq:hij}
\end{equation}
and that power system dynamics have the following property.

\noindent\begin{property} Matrices $A_i$, $B_i$, $C_i$, and $H_{ij}$ may be in general nonlinear but they are uniformly bounded in the whole state space. \label{property} \end{property}

Further discussion will be provided in section \ref{sec:robustness} to illustrate this property as well as robustness of the proposed control design.

It should be noted that equivalent network equations \eqref{eq:gi-flow} and \eqref{eq:gi-pf} are linear in system parameters. Hence, using the data locally available from WAMS, the linear equations in \eqref{eq:gi} can be solved distributively to estimate the system parameters $G_{ij}$ and $B_{ij}$. For instance, a distributed  algorithm was proposed in \cite{Azwirman2018dist} to solve linear equations and determine the system topology matrix using only local information, and this approach has successfully been applied in \cite{2018CDC} to determine a reduced-order dynamic model of power systems. In short, distributed and efficient algorithms can be developed to monitor both the system parameters and the values of $h_{ij}(y_{i1},y_{j1})$ online.
{blue}{Once again, the proposed design needs only the knowledge of upper and lower bounds on $h_{ij}(y_{i1},y_{j1})$ in order to ensure robustness.} 

\subsection{Individual Control Design}

Consider nominal subsystem \eqref{nominalsystem} and its corresponding storage function \eqref{fcnv}. It follows from \eqref{PS} that the nominal subsystem is passivity-short and $L_2$ stable if $K_i$ is designed to ensure data-driven matrix inequality \eqref{DMI}. There are many choices of self-feedback control gain $K_i$ (or gain function $K_i(x_i)$). As shown in appendix \ref{sec:modeling}, traditional generators and inverter-based energy resources have the relative degrees of two or higher, and hence the inter-area power dispatch control problem always involves passivity-short subsystems. Recall that property \ref{property} holds for power systems. Accordingly, individual subsystems can always be made passivity-short, as the same conclusion is drawn in \cite{joo2016} for any linear stabilizable systems.

From the perspective of minimizing the impact of subsystem dynamics on the overall system, an effective self-feedback control can be designed by minimizing $\epsilon_{ii}$ and maximizing $\rho_i$. That is, the following real-time optimization problem with constraints of the data-driven matrix inequality (DMI) can be solved:
\begin{equation}
\begin{split}
\min_{K_i,\epsilon_{ii},\rho_i}\;\; & [\alpha_{ii}\epsilon_{ii} - (1-\alpha_{ii})\rho_i]\\
s.t.& \;\; P_i >0, \;\; M_i (x_i) \leq 0, \;\; \epsilon_{ii},\rho_i\geq 0,
\end{split}
\label{DMI-nominal}
\end{equation}
where $\alpha_{ii}\in(0,1)$ is a design parameter. At any instant of time $t$, $x_i(t)$ becomes known as well as matrix $\overline{A}_i(x_i)$, and hence $K_i$ can be designed adaptively by using available (or previously determined) Lyapunov function $P_i>0$.

\subsection{Quantifying Passivity-Shortage Impact of Interconnections}

The following lemma, whose proof is included in appendix \ref{Appen:dmi}, provides a useful property for subsystem \eqref{generalsystem-i}. In particular, the quadratic terms $\epsilon_{ij}\|y_j\|^2$ quantify the impact of nonlinear interconnections on subsystem \eqref{generalsystem-i} in a way parallel to that of $\epsilon_{ii}\|u_i\|^2$.

\begin{lemma}\label{lm1}
	Subsystem \eqref{generalsystem-i} has the property that
\begin{equation}
\dot V_i \leq u_i^Ty_i + \frac{\epsilon_{ii}}{2}\|u_i\|^2 -\frac{\rho_i}{2}\|y_i\|^2 +  \frac{1}{2}\sum_{j\in {\cal N}_i} \epsilon_{ij}\|y_j\|^2, \label{IFP}
\end{equation}
provided that $M_i^{\prime}(x_i,y_j) \leq 0$, where
	\begin{equation}
	M_i^{\prime}  \stackrel{\triangle}{=} M_i - \sum_{j\in {\cal N}_i}\left( P_i H_{ij} C_i + C_i^T H_{ij}^T P_i-\frac{1}{\epsilon_{ij}} P_i  H_{ij}H_{ij}^TP_i \right).
	\label{DMI-Mp}
	\end{equation}
\end{lemma}

While $u_i$ can be designed to damp out inter-area oscillations, transient impacts of those oscillations must be minimized. Accordingly, the following optimization problem can be solved by using the DMI in \eqref{DMI-Mp}:
\begin{equation}
\begin{split}
\min_{\epsilon_{ij}}\;\; & \sum_{j\in{\cal N}_i}\alpha_{ij}\epsilon_{ij} \\
s.t. & \;\; P_i >0, \;\; M^{\prime}_i (x_i,y_j) \leq 0, \\
& \;\; \epsilon_{ij},\alpha_{ij}\geq 0, \;\; \sum_{j\in{\cal N}_i}\alpha_{ij}=1.
\end{split}
\label{DMI-interconnection}
\end{equation}

It is worth noting that the optimization problems in \eqref{DMI-nominal} and \eqref{DMI-interconnection} can be combined into one as:
\begin{equation}
\begin{split}
\min_{K_i,\epsilon_{ii},\epsilon_{ij},\rho_i}\;\; & \left[ \sum_{j\in {\cal N}_i \cup \{i\}}\alpha_{ij} \epsilon_{ij}- \left(1-\sum_{j\in {\cal N}_i\cup \{i\}}\alpha_{ij} \right) \rho_i\right] \\
s.t.& \;\; P_i >0, \;\; M_i^{\prime} (x_i,y_j) \leq 0, \\
& \;\; \epsilon_{ii},\;\epsilon_{ij},\rho_i\geq 0, \;\; \alpha_{ii}  + \sum_{j\in {\cal N}_i}\alpha_{ij}<1.
\end{split}
\label{LMIop}
\end{equation}
The above DMI-based optimization problem can be solved by using any of the standard techniques available to solve either LMIs or bilinear matrix inequality (BMI) optimizations \cite{VANANTWERP2000363}. It is apparent that, if $H_{ij}=0$ for any $j$, $\epsilon_{ij}\equiv 0$ is the corresponding solution.

\subsection{Communication-Enabled Wide-Area Control Design}

It follows from \eqref{eq:wa-con} that the network level cooperative control
can be written in the vector form
\begin{equation}
u=-K_c L y,
\label{ccvi1}
\end{equation}
where $K_c=\text{diag}\{k_{c_i}\}$, $S^c=\{S_{ij}^c\}$, $D=\text{diag}\{S^c\textbf{1}\}$, $L=(D-S^c)$ is the Laplacian of communication network of  wide-area control. Design of the network level control depends on properties of individual subsystems, specifically, {blue}{their impact coefficients and $L_2$ parameters are quantified by $\{\epsilon_{ii}, \cdots, \epsilon_{ij}, \cdots\}$ and $\rho_{i}$, respectively.} For convenience of expression, let us denote
 \[W= \text{diag}\{\epsilon_{ii}\},\;\Gamma=\text{diag}\{\gamma_i\}, \;\text{and }\Phi = \text{diag}\{\phi_i \},\]
where $\gamma_i$ are entries of the first left eigenvector of $L$ (that is, $\gamma^T L=0$) and \[ \phi_i={\gamma_i\rho_i - \sum_{j=1:n; \; i\in {\cal N}_j} \gamma_{j}\epsilon_{ji}}. \]
Stability of the overall system can be achieved by the choice of gain matrix $K_c$, as shown by the following theorem.
\begin{theorem}
Under inequality \eqref{IFP}, system \eqref{generalsystem} exponentially converges to the desired output consensus under cooperative control \eqref{ccvi1} provided that gain $k_{c_i}\approx k_c$ is chosen as follows:
\def\theenumi{\roman{enumi}}
\begin{enumerate}[\roman{enumi}]
	\item If $\Phi \geq 0$, then
	\begin{equation}
	0\leq k_{c}<\frac{\lambda'(\Gamma L+L \Gamma)}{\lambda_{max}(L^T W\ L) }.
	\label{eq:kci}
	\end{equation}

	\item If some $\phi_i$ are negative but both inequalities $\sum_i\phi_i>0$ and $\lambda_b^2 + 4\lambda_a\min(\phi_{i})\geq0$ holds, then
    \begin{align}
		k_c \in & \left( \frac{\lambda_b-\sqrt{\lambda_b^2 + 4\lambda_a\min(\phi_{i})}}{2\lambda_a }, \right. \nonumber \\
		 & \left. \hspace*{.2in} \frac{\lambda_b+\sqrt{\lambda_b^2 + 4\lambda_a\min(\phi_{i})}}{2\lambda_a} \right), 	\label{eq:kcii}
    \end{align}

\end{enumerate}
where  $\lambda_a=\lambda_{max}(L^T W L)$, $\lambda_b=\lambda'(\Gamma L^T + L\Gamma)$, and $\lambda'(\cdot)$ and $\lambda_{max}(\cdot)$ denote the dominant eigenvalue (the smallest non-zero) and the largest eigenvalue, respectively.
	\label{lmkcl}
\end{theorem}

{\bf Proof.}
It follows from Perron-Frobenius theorem \cite{qu2009cooperative} that, as long as Laplacian $L$ (or equivalently $S^c$) is strongly connected, its left eigenvector $\gamma = \mbox{vec}\{\gamma_i\}$ is positive. Given storage functions \eqref{fcnv} for interconnected subsystems \eqref{generalsystem-i}, choose the overall storage function as
\[ V = \sum_{i=1}^n \frac{\gamma_i}{k_{c_i}}  V_i. \]
It follows from \eqref{IFP} that
\begin{eqnarray}
\dot V & = & \sum_{i=1}^n \frac{\gamma_i}{k_{c_i}}\dot V_i \nonumber \\
& \leq& \sum_{i=1}^n \frac{\gamma_i}{k_{c_i}} y_i^T u_i +\frac{1}{2} \sum_{i=1}^n \frac{\gamma_i}{k_{c_i}}\epsilon_{ii}\|u_i\|^2 \nonumber \\
&& -\frac{1}{2}\sum_{i=1}^n \frac{\gamma_i}{k_{c_i}} \rho_i\|y_i\|^2  +\frac{1}{2}\sum_{i=1}^n\sum_{j\in{\cal N}_i} \frac{\gamma_i}{k_{c_i}} \epsilon_{ij} \|y_j\|^2  \nonumber \\
& = &  - \frac{1}{2} y^T Qy,
\label{lypsf}
\end{eqnarray}
where
\begin{equation}
Q= \Gamma L^T+L \Gamma - L^T K_c W L + \Psi,
\label{lyprcd}
\end{equation}
and
\begin{equation}
\Psi =\text{diag}\{\psi_{i}\}\stackrel{\triangle}{=}
\text{diag}\left\{\frac{\gamma_i \rho_i }{k_{c_i}} - \sum_{j=1:n; \; i\in {\cal N}_j}  \frac{\gamma_j }{k_{c_j}} \epsilon_{ji} \right\}.\label{eq:psi}
\end{equation}

Hence, the overall system is exponentially stable if matrix $Q$ is positive definite, and it has both Lyapunov stability and an output consensus if matrix $Q$ is both positive semi-definite and of rank $(n-1)$.
Should $k_{c_i}=k_c$, equation \eqref{lyprcd} reduces to
\begin{equation}
Q = -k_{c} L^T W L+ \Gamma L^T+L \Gamma  + \frac{\Phi}{k_c} .
\label{eq:optkc}
\end{equation}

It follows from \cite{qu2009cooperative} that, if $L$ is strongly connected, $(\Gamma L+L \Gamma)$ is positive semi-definite and of rank $(n-1)$, and so is $L^T K_c W L$. Therefore, $(\Gamma L+L \Gamma - k_c L^T W L)$ is positive semi-definite and of rank $(n-1)$ for all small values of $k_{c}$ satisfying \eqref{eq:kci}. Hence, stability can be concluded for the case that $\Phi\geq0$.

In the event that $\Phi\not\geq0$, some of $\phi_i$ must be negative, and stability will be established in two steps. First, note that both $(\Gamma L^T + L\Gamma)$ and $L^T W L$ and $\Phi$ are positive semi-definite and of rank $(n-1)$, in particular,  ${\bf 1}^T[-k_{c} L^T W L+(\Gamma L^T+L \Gamma )]{\bf 1} ={\bf 0} $ if and only if $x=c{\bf 1}$, where ${\bf 1}$ is the vector of $1$s. It follows that
\[ {\bf 1}^TQ{\bf 1} = \frac{1}{k_c}{\bf 1}^T\Phi{\bf 1}>0. \]
Second, it follows from \eqref{eq:optkc} that, for $x \notin \{c{\bf 1}\}$ with $c\in\Re$ and $c\not=0$,
\begin{align}
x^TQx \geq & -k_{c}\lambda_{max}( L^T W L)\|x\|^2+\lambda'(\Gamma L^T+L \Gamma )\|x\|^2 \nonumber \\
& + \frac{\min(\phi_{i})}{k_c}\|x\|^2,
\label{eq:optkcc}
\end{align}
which is positive for all $k_c$ satisfying inequality \eqref{eq:kcii}. This concludes the proof. \hfill $\triangle$

\section{Implementation and Its Robustness}

\begin{figure}[ht]
	\centering \includegraphics[height=3cm]{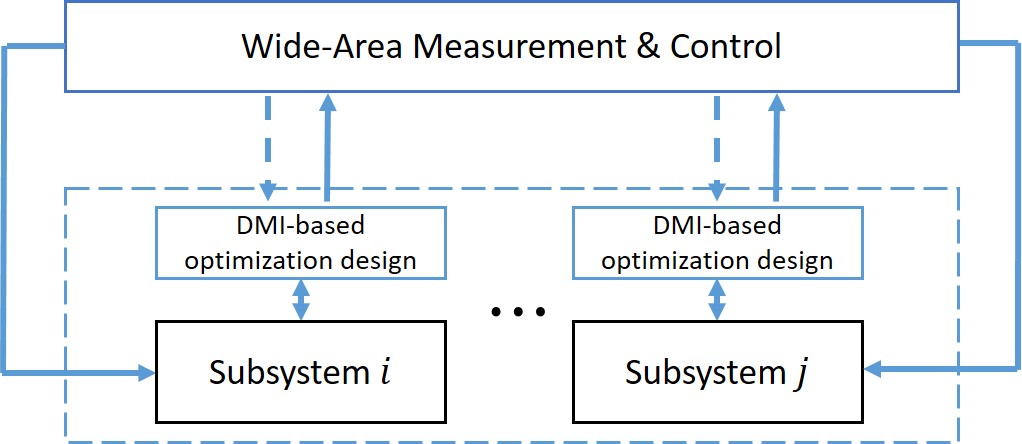}
	\caption{Overall architecture of the proposed control}
	\label {OAC}
\end{figure}

The proposed design yields a two-level control implementation: a lower-level control involving feedback gain matrix $K_i$ for passivity shortage and $L_2$ stability of the subsystems, and a higher-level control enabled by WAMS. The control architecture is shown in Fig. \ref{OAC}, and the details of communication topology design, control algorithm implementation and robustness are illustrated in the subsequent subsections.

\subsection{Communication Topology Design \label{sec:comm}}

Implementation of wide-area cooperative control \eqref{eq:wa-con} or \eqref{ccvi1} involves the choice of communication matrix $S^c$, a mathematical abstraction of WAMS. The simplest choice of $S^c$ is
\[ S^c={\bf 1}{\bf 1}^T, \;\;\; \mbox{or} \;\;\; S^c_{ij}\equiv 1, \]
which is all to all communication.

Wide-area control $u_i$ in \eqref{eq:wa-con} is in the form of consensus law \cite{qu2009cooperative}, and it has the property that, should subsystems $i$ and $j$ are {\it coherent} (a concept established in power systems in \cite{sastry1981coherency} or, equivalently, in the sense that $(y_i-y_j)\approx 0$), the corresponding control contribution is approximately zero. In other words, WAMS-enabled control \eqref{eq:wa-con} aims specifically at damping out inter-area oscillations, and communication neighboring set ${\cal N}_{c_i}$ should be chosen to exchange information among incoherent groups of subsystems (i.e., subsystems that are geographically apart). That is, the sparsest communication matrix is as follow: given $\mu$ coherent groups of generators and for any $1\leq l_1\not= l_2\leq \mu$, $S^c_{ij}=0$ for all $i\in {\cal C}_{l_1}$ and $j\in {\cal C}_{l_2}$ except for one pair $(i^*,j^*)$ such that $S^c_{i^*j^*}=0$ with $i^*\in {\cal C}_{l_1}$ and $j^*\in {\cal C}_{l_2}$, where ${\cal C}_l$ is the index set of the $l$th coherent group, and $\cup_{l=1}^{\mu} {\cal C}_j ={\cal N}$.

Any topology more dense than the sparsest would work for the proposed wide-area control. Under the topology of all-to-all communication, system parameters $G_{ij}$ and $B_{ij}$ can be solved at all the sites. Under all other possible topologies, system parameters can be estimated either by using distributed algorithm explained in section \ref{sec:H} or by the dispatch control center (which collects all critical information of the overall system).

\subsection{Implementation of Data-Driven Control \label{DDC2}}

The data used in the proposed design contains:
\begin{description}
    \item[Step 1: ] $\;$Local model and data at individual systems: matrices $A_i$, $B_i$, and $C_{i}$; local state $x_i$ and output $y_i$; and the outcome of individual feedback gain design is $P_i$, $K_i$, $\epsilon_{ii}$, and $\rho_i$;
    \item[Step 2: ] $\;$Wide-area data-driven interconnection model and its impact: outputs $y=\mbox{vec}\{y_i\}$; estimates of system parameters $G_{ij}$, $B_{ij}$, and values $H_{ij}$; determination of impact measures $\epsilon_{ij}$ through optimization;
    \item[Step 3: ] $\;$\noindent Wide-area cooperative control: determination of $k_c$ by applying theorem \ref{lmkcl}.
\end{description}
Should matrices $A_i$, $B_i$, and $C_{i}$ be constant, step 1 needs to be done only once; otherwise, step 1 needs to be dynamically computed if the matrices are state dependent. In general, step 2 needs to be dynamically calculated as wide-area data comes in. Steps 1 and 2 can be integrated into one as illustrated by \eqref{LMIop}. Step 3 needs to be repeated if $\epsilon_{ij}$ change noticeably.

It follows from \eqref{eq:hij} that the changes in matrices $H_{ij}$ are due to their elements $h_{ij}(y_{i1},y_{j1})$.  As will be discussed in section \ref{sec:robustness}, the change of $h_{ij}$ over time is often small and relatively slow. As a result, wide-area data-driven computation can be held off until the cumulative change of $h_{ij}$ has exceeded certain threshold $c_{T_i}$.  Specifically, let us define the following measure:
\begin{equation}
\overline{h}_i(t) \stackrel{\triangle}{=} \sum_{j \in {\cal N}_i} | h_{ij}(y_{i1}(t),y_{j1}(t))|,
\label{eq:hba}
\end{equation}
then step 2 is ignored for the $i$th generator over time interval $[t-\delta t, \; t]$ if
\begin{equation} |\overline{h}_i(t) - \overline{h}_i(t-\delta t)| < c_{T_i}. \label{eq:cT}
\end{equation}

In summary, the proposed data-driven control is implemented using algorithm \ref{ddp}:
\begin{algorithm}
\caption{Computational algorithm of data-driven control}\label{ddp}
\begin{algorithmic}[1]
\State At time $t_i$, update $ H_{ij}(t_i)$ and check condition \eqref{eq:cT}:

\quad if \eqref{eq:cT} holds, \textbf{exit};

\quad else, continue.

\State Initialization: $K_i$, $\epsilon_{ii}$, $\epsilon_{ij}$, $\rho_i$, and $P_i$.

\State Update $ H_{ij}(t_i)$ by wide-area data (also update $A_i$, $B_i$, and $C_{i}$ if they are state-dependent).

\BState Perform {\it DMI optimization process} at time $t_i$ until it converges:

\State \quad\quad\quad Update $M'_i$ according to \eqref{DMI-Mp};
\State \quad\quad\quad Solve bilinear problem \eqref{LMIop}.

\State Update $W$, $\Phi$, and choose $k_c$  according to theorem \ref{lmkcl}.
\State \textbf{goto} $t_{i+1}$.
\end{algorithmic}
\end{algorithm}
\subsection{{blue}{Time Delay and Robustness Analysis}\label{sec:robustness}}

\begin{figure}
	{\centering \includegraphics[width=9cm]{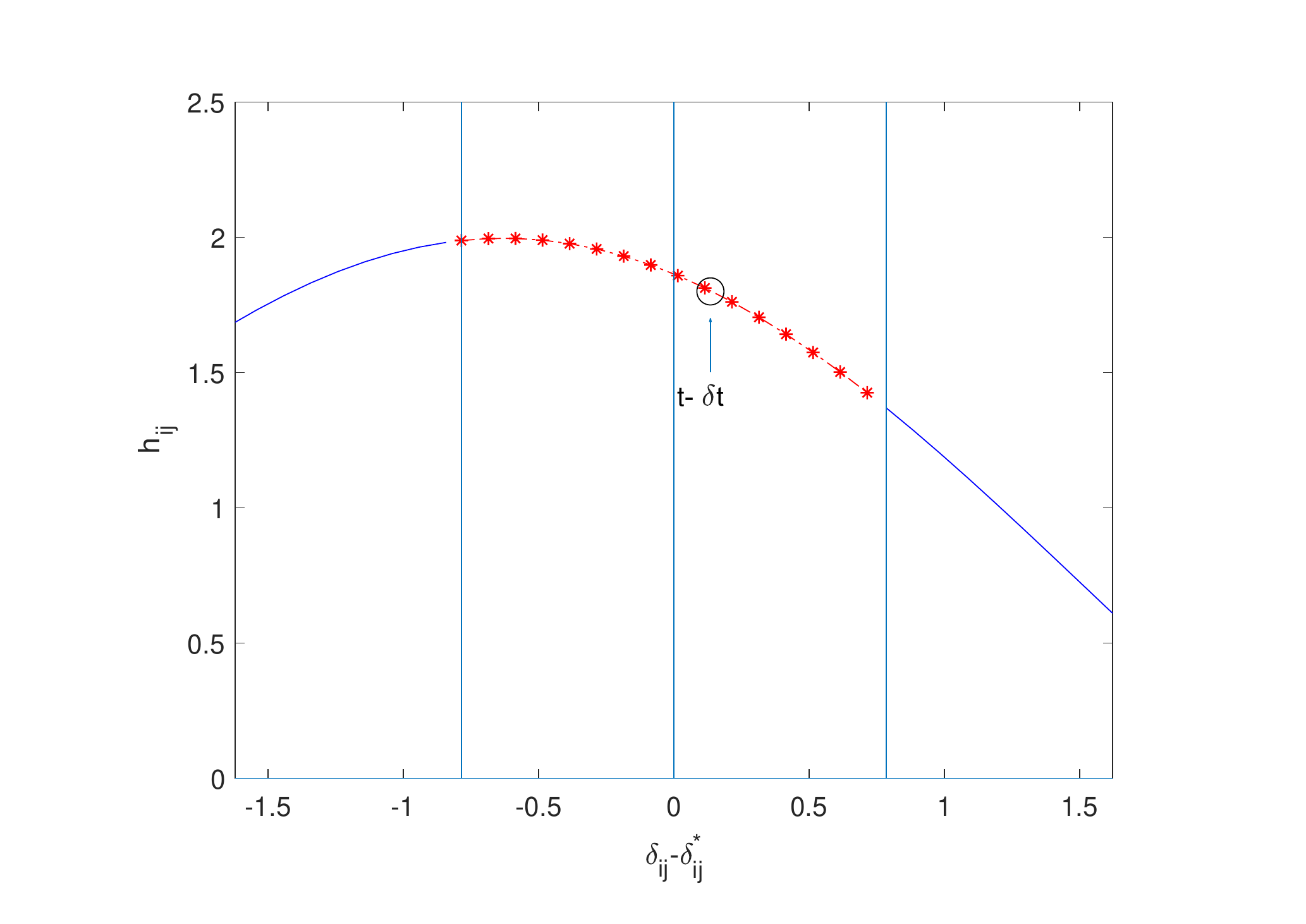}}
	\caption{Range of $h_{ij}$ values with respect to  $(\delta_{ij}-\delta_{ij}^*)$}
	\label{fig:hij_p}
\end{figure}

It is straightforward to show using \eqref{eq:hij} that, while $h_{ij}(y_{i1},y_{j1})$ are nonlinear, their values are uniformly bounded from above and below. Indeed, upon determining system parameters $G_{ij}$ and $B_{ij}$, the ranges of $h_{ij}(y_{i1},y_{j1})$ can easily be found for a wide operational range of $\delta_i$ and its equilibrium $\delta_i^*$. Fig. \ref{fig:hij_p} is an illustration of the range of $h_{ij}$, which is drawn based on the parameters from appendix \ref{Appen:netpar}. The deviation of $h_{ij}$ as shown by the middle part of the curve is quite small when the system has a low-frequency oscillation (i.e. $(y_{i1}-y_{j1})$ fluctuates around zero).

It follows from Property \ref{property} that the optimization problem \eqref{DMI-interconnection} remains to be solvable only with the ranges of $H_{ij}$ rather than their precise values.  As such, a solution to the optimization problem \eqref{LMIop} could be found without the exact knowledge of $y_j$. {blue}{In case that $y_j$ was available at $(t-\delta t)$ rather than current time instance $t$ because of communication delay or interruption,  lets define the following error of $h_{ij}$ to analyze robustness:}
\[\Delta h_{ij}=h_{ij}(t)-h_{ij}(t-\delta t),\]
and hence that of $H_{ij}$ can be written accordingly as
\[\Delta H_{ij}\stackrel{\triangle}{=}H_{ij}(t)-H_{ij}(t-\delta t) =\frac{\Delta h_{ij}}{h_{ij}(t-\delta t)}H_{ij}(t-\delta t). \]
Therefore, by substituting $(H_{ij}(t-\delta t)+\Delta H_{ij})$ into \eqref{LMIop}, we have the following DMI matrix at $t$
\begin{equation*}
\begin{split}
 M'(t)=&M'(t-\delta t)
 +\left(\frac{\Delta h_{ij}}{h_{ij}(t-\delta t)}\right)^2N_{ij}(t-\delta t)
 \\&+\frac{\Delta h_{ij}(t-\delta t)}{h_{ij}(t-\delta t)}N'_{ij}(t-\delta t),
\end{split}
\end{equation*}
where $N_{ij}(t-\delta t)=\frac{1}{\epsilon_{ij}h_{ij}(t-\delta t)} P_i H_{ij}(t-\delta t)H_{ij}^T(t-\delta t) P_i$, and
\begin{equation*}
\begin{split}
N'_{ij}(t-\delta t)=&\frac{1}{\epsilon_{ij}h_{ij}(t-\delta t)}\left[2P_i H_{ij}(t-\delta t)  H_{ij}^T(t-\delta t) P_i \right.\\ &\left.-\epsilon_{ij}(P_i  H_{ij}(t-\delta t) C_i + C_i^T  H_{ij}^T(t-\delta t) P_i)\right].
\end{split}
\end{equation*}
Given  $M'(t-\delta t)\leq 0$, it can be observed that $M'(t)\leq 0$ holds if
\begin{equation*}
\begin{split}
\|\Delta h_{ij}\|^2&\|N_{ij}(t-\delta t)\| + \|\Delta h_{ij}\|\|N'_{ij}(t-\delta t)\| \\&\leq \lambda_{min} \left(M'(t-\delta t)\right),
\end{split}
\end{equation*}
which is equivalent to
\begin{equation}
\begin{split}
\|\Delta& h_{ij}\|\leq \frac{1}{2\|N_{ij}(t-\delta t)\|}\left[-\|N'_{ij}(t-\delta t)\|\right.\\
&\left.+\sqrt{\|N'_{ij}(t-\delta t)\|^2+4\|N_{ij}(t-\delta t)\| \lambda_{min} \left(M'(t-\delta t)\right)}\right]
\end{split}
\end{equation}
where $\lambda_{min}(\cdot)$ denotes the smallest eigenvalue. {blue}{The above bound can be integrated into $c_{T_i}$ in \eqref{eq:cT}, the condition for algorithm \ref{ddp}. Based on the fact that deviation $\Delta h_{ij}$ is usually small for low-frequency oscillations, robustness of algorithm \ref{ddp} is assured, and hence each individual subsystem is guaranteed to be passivity-short and $L_2$ stable in the presence of delay $\delta t$. Using the passivity shortage framework, the overall system stability under consensus-based cooperative control (16) through delayed communication network remains to be stabilizing albeit its performance is degraded graciously as delay increases. Due to space limitation, further analysis and analytical proof are omitted here but the readers are referred to \cite{qu2009cooperative,venka2018pass} for the detailed analysis on stability of interconnecting  passivity-short systems with significant communication delays.}

\def\theenumi{\roman{enumi}}

\section{Simulation Results}
\begin{figure}
 \includegraphics[width=8.5cm]{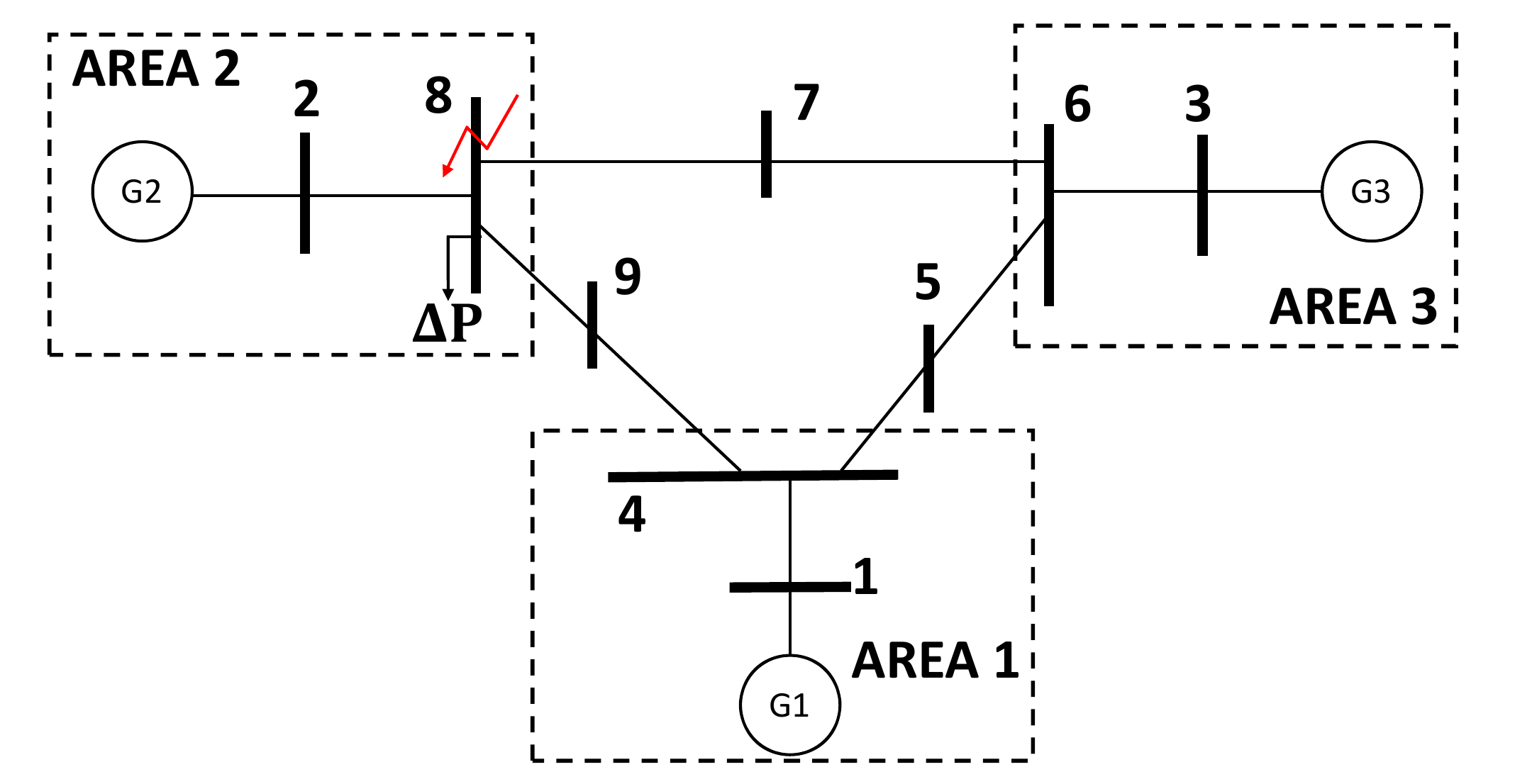}
 \caption{A three-area power system: IEEE 9-bus system}
 \label{3-area}
 \end{figure}

{blue}{In this section, performance of the proposed control is illustrated using a three-area test system modeled by the standard IEEE 9-bus system shown in Fig. \ref{3-area}. Each of the areas is represented by one aggregated generator. The detailed parameters of the test system and its aggregated generator models are listed in appendix \ref{Appen:netpar}.}

{blue}{All the simulation results are obtained using the following setting: a short-circuit fault happens on bus 8 at $t=2.0$s and is removed at $t=2.1s$, and a load change of $\Delta P_{load} = 1.\; p.u.$ occurs simultaneously at $t=2.1s$ in area 2. In this setting, there are a large disturbance and a consequent change of operating condition.}

{blue}{Performance of the following three wide-area controls are compared under the same simulation setting: I) the proposed DMI control, II) the LMI-designed control \cite{Chilali1996Hinfi}, and III) a traditional control of typical gain choice (constant gains). For case III,  the local gain matrix is chosen to be a droop gain, and the wide area control is AGC whose gain matrix is an integral control, that is, in \eqref{eq:con} and \eqref{eq:Ui},
\[K_i=\left[0, k_i, 0,0,0\right],\;\; K_i^I=\left[0, k_i^I, 0,0,0\right].\]
The specific gain values in case III are set to be $k_i=30$ and $k_i^I=0.3$, as suggested in \cite{kundur1994power}, for all three aggregated generators. The control gains in case II are solved using the LMI method \cite{Chilali1996Hinfi} based on the operating condition. The control gains in case I are updated using the proposed DMI optimization procedure, and network-level control gain $k_c$ is solved subsequently. }

\begin{figure}[ht]
	\includegraphics[width=9cm]{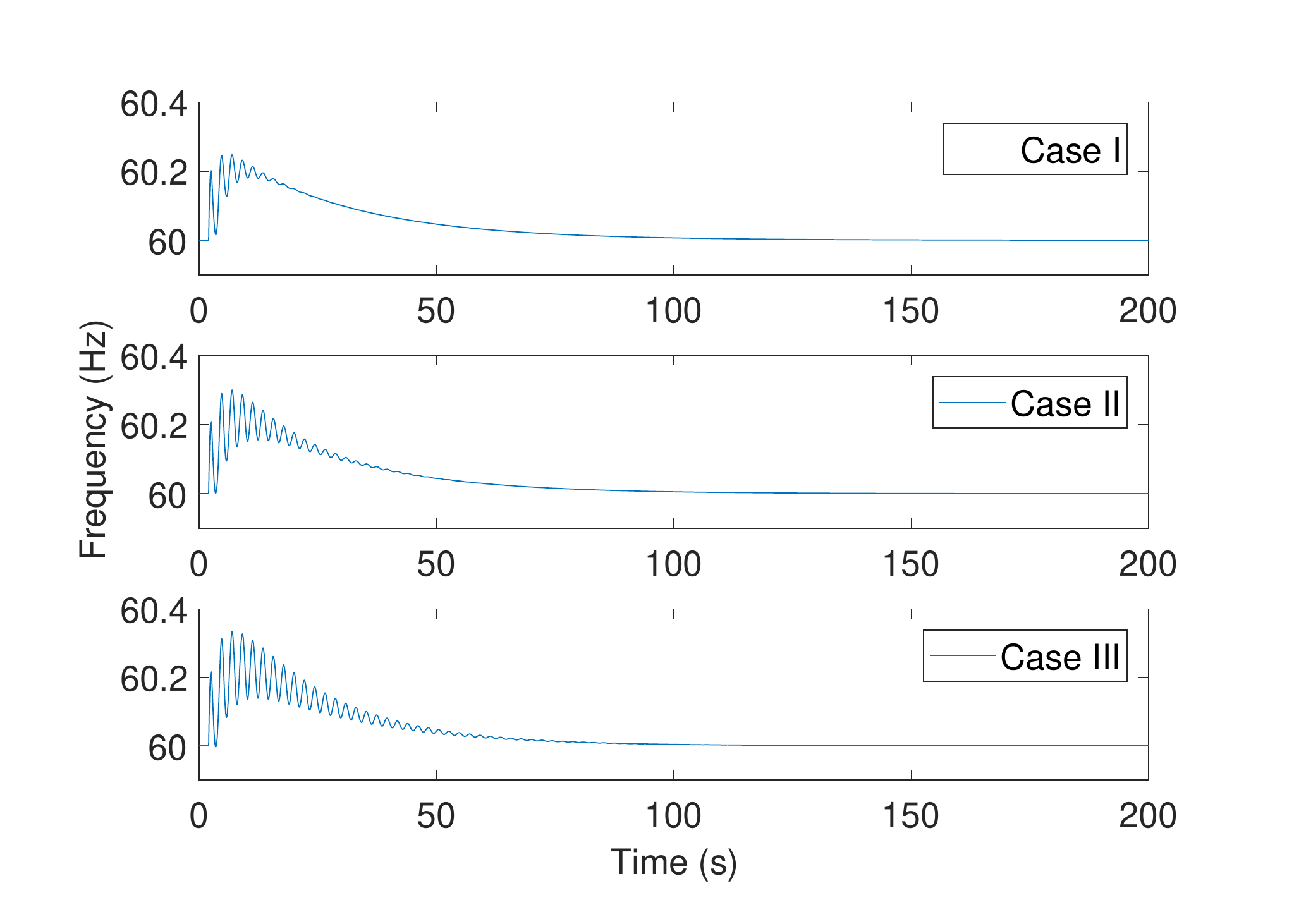}
	\caption{Trajectories of $\omega_3$: the proposed control (case I), LMI control (case II), and traditional control (case III)}
	\label{p2}
\end{figure}

\begin{figure}[ht]
	\includegraphics[width=9cm]{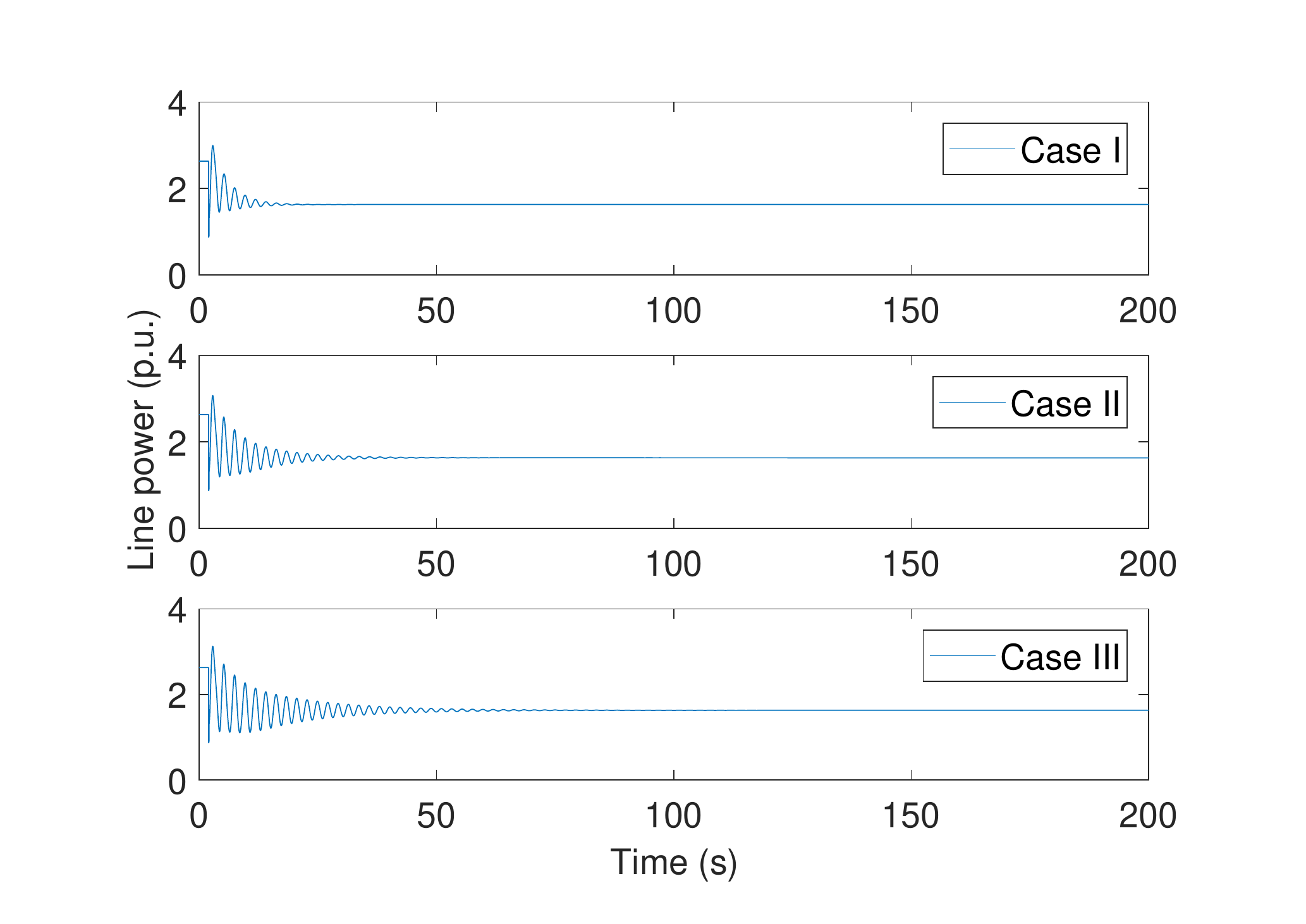}	
	\caption{Trajectories of power on line 8-2: the proposed control (case I), LMI control (case II), and traditional control (case III)}
	\label{pe}
\end{figure}

{blue}{
The system responses under the three control strategies are illustrated by both the frequency and a line power (line 8-2), and the results are shown in Fig. \ref{p2} and Fig. \ref{pe}, respectively. The results show that system frequency increases as a result of the load decrease in area 2, after sustained oscillations. The oscillation frequency is about 0.4Hz. Fig. \ref{p2} also shows that system frequency gradually goes back to nominal frequency in all the cases.
}
\begin{figure}[ht]
	\includegraphics[width=9cm]{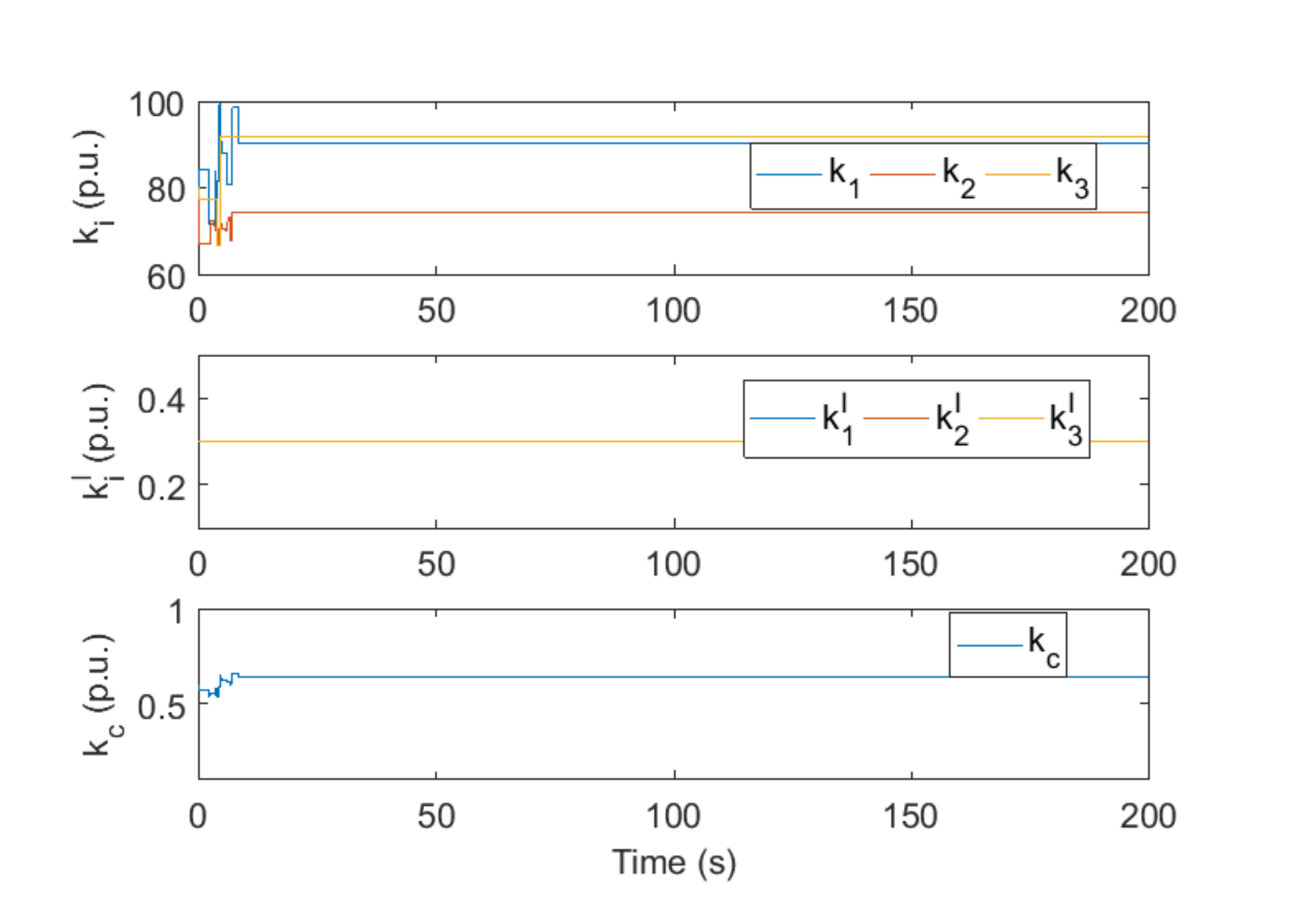}
	\caption{Adaptation over time of control gains in the proposed DMI design: $k_i$, $k_i^I$ and $k_c$.}
	\label{fig:gains}
\end{figure}

{blue}{The representative simulation results show that the proposed DMI design method is more effective in damping out the oscillations caused by the disturbances. Case III has the least damping because the typical AGC is not effective in oscillation suppression. The performance in case II is somewhat better as the control gains are optimized but only with respect to one loading condition. Changes of the operating condition are not considered in either cases II or III. In case I, the overall system stability is monitored using matrix inequalities, the control gains are adaptively updated, and hence the best performance is achieved. Fig. \ref{fig:gains} shows the time evolution of the adaptive control gains in the proposed data-driven control. From both analytical design and the simulations, we learn that $k_i$ in \eqref{LMIop} is far more effective than $k_i^I$ (as $k_i^I$ has to be small).}

\begin{figure}[ht]
	\includegraphics[width=9cm]{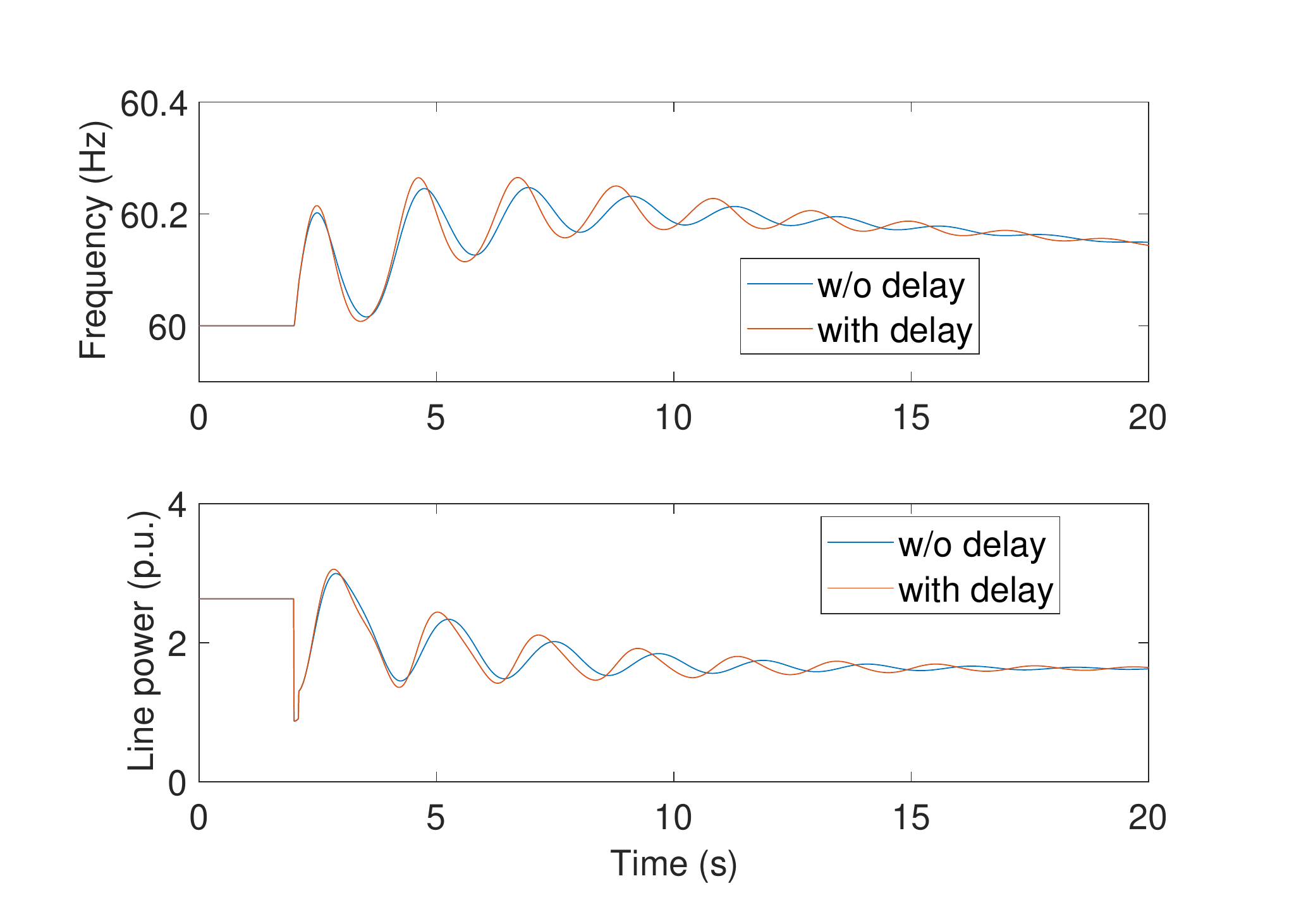}
	\caption{Damping performance of the proposed DMI control: Without or with communication delay.}
	\label{fig:delay}
\end{figure}

{blue}{To illustrate performance of the proposed DMI control with delayed communication, the simulation is repeated with delay $\delta t =  200$ms (the same value used in \cite{wang2012wide}). The results in Fig. \ref{fig:delay} show that the proposed DMI control remains quite effective and performance degradation is in tune with the delay. }
\section{Conclusion and Discussion}
A novel systematic approach for damping out inter-area low-frequency oscillations of power systems is presented. By taking advantage of the fact that all synchronous generators are passivity-short and $L_2$ stable, the passivity shortage framework is utilized to modularly design a two-level control. The stability of the overall system is investigated as the interconnection of subsystems, local control, and wide-area control. First, using WAMS data, a DMI algorithm is formulated to minimize the impact of individual control of subsystems and their interconnections. Then, a high-level control algorithm is designed and analyzed to ensure the overall system stability. Simulation on a standard test system proves the efficacy of the proposed method. Implementation of a passivity-short design results in a closed-form representation in which overall system stability can be guaranteed.

{blue}{The proposed data-driven control is robust with respect to network parameter changes and communication delays or interruptions. Its implementation only requires computation of matrix inequalities, hence the proposed control is scalable to large-scale systems. Nonetheless, it is natural to model and control groups of coherent generators together. Future research will be pursued to address such technical issues as joint design of communication, on-line model reduction, and wide-area control. Such a comprehensive solution is what is needed for WAMS and EMS. }

\ifCLASSOPTIONcaptionsoff
  \newpage
\fi




\bibliographystyle{IEEEtran}
\bibliography{IEEEabrv,wide_jrnl}

\appendices
\renewcommand{\theequation}{\thesection.\arabic{equation}}
\setcounter{equation}{0}

\section{State-Dependent Affine Modeling of Power Systems} \label{sec:modeling}
Consider a power system that consists of $n$ generators. If the $i$th generator is conventional, its  dynamic equations are described by the following swing equations:
\begin{equation}
\dot{\delta_i}=\omega_i,\;\;
M_i\dot \omega_i=P_{m_i} - P_{g_i} - D_i  \omega_i.
 \label{eq:sw}
\end{equation}
where $M_{i}$ is the inertia, $P_{m_i}$ is the prime power, $P_{g_i}$ is the electrical power, $D_{i}$ is the damping constant, $\delta_i$ is the rotor angle, and $\omega_i$ is the frequency derivation (away from the synchronous frequency $\omega_0$). Should the generator have a prime mover, the mechanical power $P_{m_i}$ is the output of a second-order turbine/governor model \cite{Walker1989Intg}:
\begin{equation}
\tau_{i_1} \dot P_{m_i} =  Y_{g_i} - P_{m_i},\;\;
\tau_{i_2} \dot Y_{g_i} =  U_i - Y_{g_i},
\label{eq:tg}
\end{equation}
where $Y_{g_i}$ denotes the prime mover input (e.g. the gate position of the turbine), $\tau_{i_1}$ and $\tau_{i_2}$ are time constants, and ${U}_i$ is the control input.

\begin{figure}
\includegraphics[width=8.5cm]{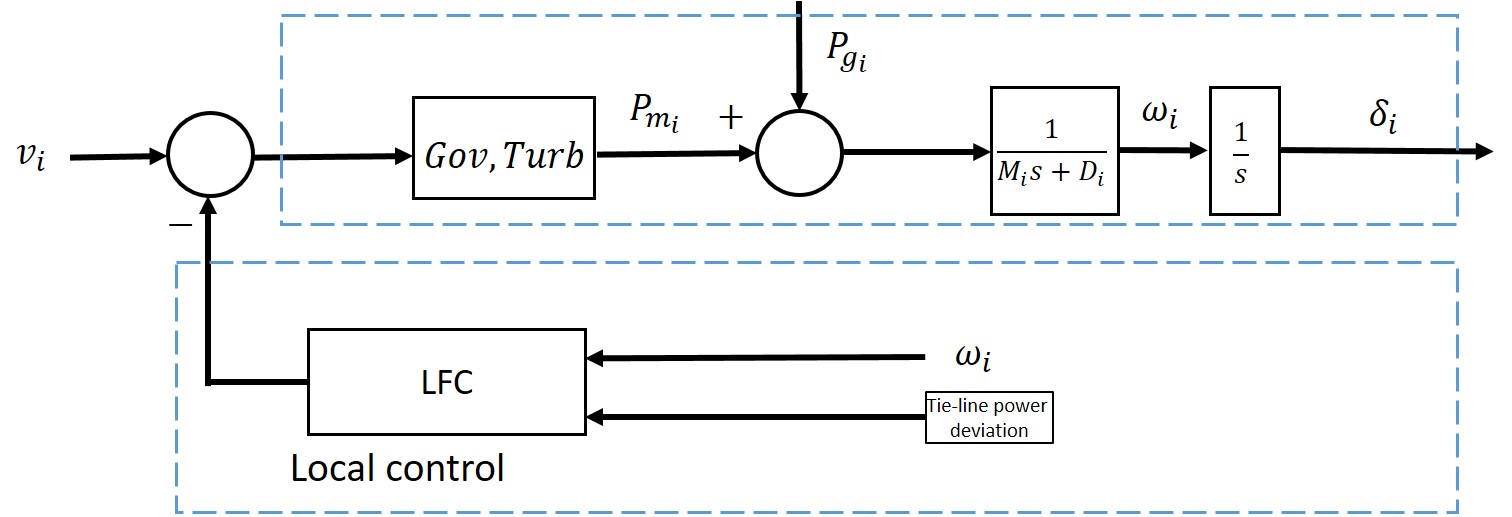}
\caption{Block diagram of droop control}
\label{local}
\end{figure}

\begin{figure}
    \centering
    \includegraphics[width=6cm]{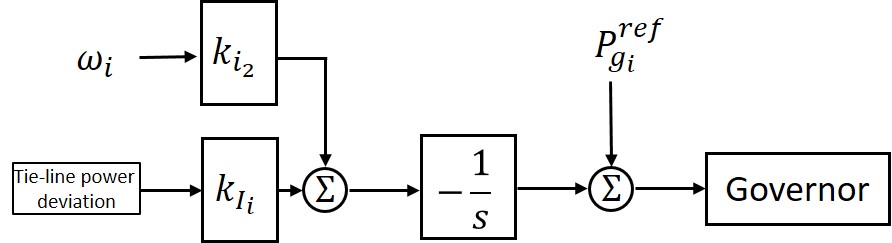}
    \caption{Block diagram of automatic generator control (AGC)}
    \label{fig:AGC}
\end{figure}

Traditionally, control ${U}_i$ consists of a local frequency control (LFC) and an automatic generator control (AGC). As shown in Fig. \ref{local} and Fig. \ref{fig:AGC}, LFC is a droop control that is a special case of the proposed feedback control $v_i=-K_ix_i + u_i$, and AGC is an integral control of the local feedback. Hence, control ${U}_i$ is in general of the form
\begin{equation}
{U}_i = v_i + \alpha_i + P_{g_i}^{ref},\;\;\; \dot{\alpha}_i = -K_i^I x_i,
\label{eq:Ui}
\end{equation}
where $x_i$ is the local state to be defined, $\alpha_i$ is the integral control variable, $P_{g_i}^{ref}$ is the desired set point, $K_i^I$ is an integral gain row vector (of small values), $v_i=-K_ix_i + u_i$, and $u_i$ is the higher-level control to be designed in the form of \eqref{eq:wa-con}.

It follows from \eqref{eq:gi}, \eqref{eq:sw}, \eqref{eq:tg}, and \eqref{eq:Ui} that the equilibrium of the overall system with $x_i^*=0$ is
\begin{subequations}
\begin{eqnarray}
P_{m_i}^* & = & P_{g_i}^* \\
Y_{g_i}^* & = & P_{g_i}^* \\
P_{g_i}^* & = & P_{g_i}^{ref} + \alpha_i^* \\
P_{g_i}^* & = &  \sum_{j \in {\cal N}} E_{i}E_{j}(G_{ij} \cos{\delta_{ij}^*}+B_{ij}\sin{\delta_{ij}^*}) \\
Q_{g_i}^* &  = & \sum_{j \in {\cal N}} E_{i}E_{j}(G_{ij} \sin{\delta_{ij}^*}-B_{ij}\cos{\delta_{ij}^*}). \hspace*{.2in}
\end{eqnarray}\label{eq:equilibrium}
\end{subequations}\noindent

In the above equations, $\delta_{ij}^{*}=\delta_i^*-\delta_j^*$ represents the final angle differences whose values are determined by the AGC signals \cite{kundur1994power} applied at each of the generators. AGC is a fundamental function of the power system control center, which adjusts the outputs of all major plants to compensate the frequency and load changes.

For a generator described by model \eqref{eq:sw} and \eqref{eq:tg}, let us choose its state and output vectors as
\begin{eqnarray}
 x_i=  \begin{bmatrix} x_{i1} \\   x_{i2} \\  x_{i3} \\  x_{i4} \\  x_{i5}  \end{bmatrix}
 = \begin{bmatrix} \delta_{i}-\delta^{*}_{i} \\   \omega_i \\ P_{m_i}-P_{g_i}^* \\ Y_{g_i}-P_{g_i}^*\\  \alpha_{i}-\alpha^{*}_{i}  \end{bmatrix},
\hspace*{.2in}
y_i  = \begin{bmatrix} y_{i1} \\   y_{i2} \end{bmatrix}
 =  \begin{bmatrix} x_{i1} \\ x_{i2} \end{bmatrix}. \label{eq:xi}
\end{eqnarray}
\vspace{0.1in}
It follows from \eqref{eq:gi} and \eqref{eq:xi} that
\begin{equation}
P_{g_i} = P^{*}_{g_i}+\sum_{j\in{\cal N} }h_{ij}(y_{i1},y_{j1}) \times (y_{i1}-y_{j1}),
\label{eq:parl}
\end{equation}
where $h_{ij}(y_{i1},y_{j1})$ is defined by \eqref{eq:hij}.

It is straightforward to show that, under the definition of $x_i$ in \eqref{eq:xi}, dynamic equations \eqref{eq:sw} and \eqref{eq:tg} together with control \eqref{eq:Ui} and load flow equations \eqref{eq:gi} are mapped into \eqref{generalsystem} with
\begin{equation}
\begin{split}
\hspace*{-0.15in} A_i = \left[ \begin{array}{ccccc} 0&1&0&0&0
	\\ 0  & -\frac{D_i}{M_i} &\frac{1}{M_i}&0&0\\
		0 &0 &-\frac{1}{\tau_{i_1}}&\frac{1}{\tau_{i_1}}&0\\
		0 & 0 &0&-\frac{1}{\tau_{i_2}}&\frac{1}{\tau_{i_2}}\\ \hdashline
		&  &-K_i^I &&\end{array} \right],\;
B_i = \begin{bmatrix}0\\0\\0\\ \frac{1}{\tau_{i_2}} \\0\end{bmatrix}, \\
\hspace*{-0.2in} C_i=\begin{bmatrix}1&0&0&0&0\\0&1& 0 &0&0\end{bmatrix}, \;\;\; H_{ij}(y_i,y_j)=\begin{bmatrix}0&0 \\ \frac{h_{ij}}{M_i}& 0  \\0&0 \\ 0 & 0\\0&0 \end{bmatrix}.
\end{split}
\label{eq:ABCH}
\end{equation}

For inverter-based renewable generation or distributed energy resources, there are at least two options to derive their dynamic equations and apply the proposed design. One is to use their native dynamics to derive their equations, as was done in \cite{Xin2011PS}. The other option is to introduce a layer of generator emulation control \cite{2011ICPE,2011IASA} or equivalently the so-called virtual synchronous generator \cite{shintai2014oscillation}. In the latter case, inverter-based energy sources have the same dynamic performance as synchronous machines, which is adopted in this paper for simplicity of technical presentation. Accordingly, for inverter-based generators, dynamic equations in \eqref{eq:sw} are used, and the corresponding state and matrices are given as follows:
\begin{eqnarray}
x_i=  \begin{bmatrix} x_{i1} \\   x_{i2} \\  x_{i3}   \end{bmatrix}
= \begin{bmatrix} \delta_{i}-\delta^{*}_{i} \\   \omega_i \\  \alpha_{i}-\alpha^{*}_{i}  \end{bmatrix},
\hspace*{.2in}
y_i  = \begin{bmatrix} y_{i1} \\   y_{i2} \end{bmatrix}
=  \begin{bmatrix} x_{i1} \\ x_{i2} \end{bmatrix}, \label{eq:xi_der}
\end{eqnarray}
and
\begin{equation}
\begin{split}
\hspace*{-0.2in} A_i = \left[ \begin{array}{ccc}0&1&0
\\ 0 & -\frac{D_i}{M_i} &\frac{1}{M_i}\\\hdashline
&  -K_i^I & \end{array} \right],\;\;\;
B_i = \begin{bmatrix} 0\\ \frac{1}{M_{i}}\\0 \end{bmatrix},
\\
\hspace*{-0.2in} C_i=\begin{bmatrix}1&0&0\\0&1& 0 \end{bmatrix}, \;\;\;
H_{ij}=\begin{bmatrix}0&0 \\ \frac{h_{ij}}{M_i}& 0  \\ 0 & 0 \end{bmatrix}.
\end{split}
\label{eq:ABCH_der}
\end{equation}
It is worth noting that generator dynamics may be subject to such nonlinearity as saturation. Correspondingly, those nonlinearities can be introduced and matrix representations \eqref{eq:ABCH} and \eqref{eq:ABCH_der} become state dependent.

{blue}{It should be noted that large-scale power systems often consist of distinct geographical regions and their coherent groups of physical generators. Rather than designing wide-area controls for all physical generators, it is both advantageous and customary to develop an aggregated model for each of the areas. Indeed, the aggregate model can also be expressed in the form of \eqref{eq:ABCH} or \eqref{eq:ABCH_der}, and they can be determined using one of model reduction algorithms in \cite{chakrabortty2011measurement,2018CDC,Scarciotti2017Reduc}).  Hence, in the paper, model \eqref{generalsystem} is used to represent either an individual generator or a group of coherent generators.  }

\section{Proof of Lemma \ref{lm1} \label{Appen:dmi}}
\setcounter{equation}{0}
It follows from subsystem \eqref{generalsystem-i} and storage function \eqref{fcnv} that
\begin{eqnarray*}
\dot V_i & = &  u_i^Ty_i + \frac{\epsilon_{ii}}{2}\|u_i\|^2 -\frac{\rho_i}{2}\|y_i\|^2 + \sum_{j\in {\cal N}_i} \frac{\epsilon_{ij}}{2}\|y_j\|^2 \\
&& +
\begin{bmatrix}
	x_i^T & y_j^T & u_i^T	
	\end{bmatrix} \overline{M}_i(x_i,y_j)\begin{bmatrix}
	x_i \\ y_j \\u_i	
	\end{bmatrix}.
\end{eqnarray*}

Inequality \eqref{IFP} can be concluded by using the above equation together with DMI (\ref{DMI-i}).

Matrix $\overline{M}_i$ in (\ref{DMI-i}) has the special structure that it has four sub-blocks as
\[ \overline{M}_i =  \begin{bmatrix}
\overline{M}_{i,11} & \overline{M}_{i,12} \\
\overline{M}_{i,12}^T & \overline{M}_{i,22}
\end{bmatrix} \]
and the lower right block $\overline{M}_{i,22}$ is diagonal and negative definite. It follows from Schur complement lemma \cite{vanantwerp2000tutorial} that $\overline  M_i \leq 0$ is equivalent to $\overline{M}_{i,22}< 0$ and
\begin{equation*}
M_i^{\prime} \stackrel{\triangle}{=}\overline{M}_{i,11}-  \overline{M}_{i,12} \overline{M}_{i,22}^{-1} \overline{M}_{i,12}^T <0.
\end{equation*}
It is straightforward to show algebraically that $M_i^{\prime}$ is given by \eqref{DMI-Mp}. clearly, the dimension of matrix $M_i^{\prime}$ is much lower than that of matrix $\overline{M}_i$.

\newpage
\begin{strip}
	\hfill
	\begin{equation}
	\overline{M}_i(x_i,y_j)=
	\left[ \begin{array}{c:cccc}
	\begin{split}
	\overline{A}_i^T(x_i) P_i + P_i \overline{A}_i(x_i) + \rho_i C_i^T C_i\\ -\sum_{j\in {\cal N}_i}( P_i H_{ij} C_i + C_i^T H_{ij}^T P_i)\end{split}
	&\cdots& P_i H_{ij}(y_i,y_j)  & \cdots &  P_iB_i-C_i^T    \\ \hdashline
	\vdots & \vdots & 0 &\vdots &  0  \\
	H_{ij}^T(y_i,y_j) P_i  & 0 & -\epsilon_{ij}   I & 0 & 0 \\
	\vdots & \vdots & 0& \vdots &0  \\
	B_i^TP_i-C_i  & 0& 0  & 0  & -\epsilon_{ii}I
	\end{array} \right] \leq 0.\label{DMI-i}
	\end{equation}
	\hfill
\end{strip}

\section{Test System Data} \label{Appen:netpar}
The network parameters are shown in Table \ref{NetPar}. The parameters of aggregated generators are shown in Table \ref{GenPar}. All parameters are per unit values.
\begin{table}[ht]
	\centering
    \caption{Network parameters of the IEEE test system}
	\begin{tabular}{c c c c c}
        \hline
 Bus 1 &Bus 2  & R& X & B \\
    \hline
1 &	4  &0   & 0.0576   & 0    \\
4 & 5 &0.017   & 0.092  & 0.158    \\
5& 6 &0.039   & 0.17 & 0.358    \\
3& 6 &0  & 0.0586   & 0    \\
6 & 7  &0.0119 & 0.1008 & 0.209  \\
7 & 8  &0.0085 & 0.072  & 0.149  \\
8& 2  &0.00   & 0.0625 & 0    \\
8& 9  &0.032   & 0.161 & 0.306    \\
9& 4  &0.01   & 0.085 & 0.176    \\

    \hline \label{tb:lns}
	\end{tabular}
    \label{NetPar}
\end{table}
\begin{table}[ht]
	\centering
    \caption{Generator parameters of the IEEE test system}
	\begin{tabular}{ c c c c c c}
        \hline
 Gen. No.  & $M_i$& $D_i$& $X_d'$  & $\tau_{i_1}$& $\tau_{i_2}$ \\
    \hline
1     &470   & 0.1&0.0014 & 0.03 & 0.01  \\
2   &130   & 0.1  &0.0023& 0.03 & 0.01  \\
3   &62   & 0.1&0.0029  & 0.03 & 0.01  \\

    \hline
	\end{tabular}
    \label{GenPar}
\end{table}

\begin{IEEEbiography} [{\includegraphics[width=1in,height=1.25in,clip,keepaspectratio]{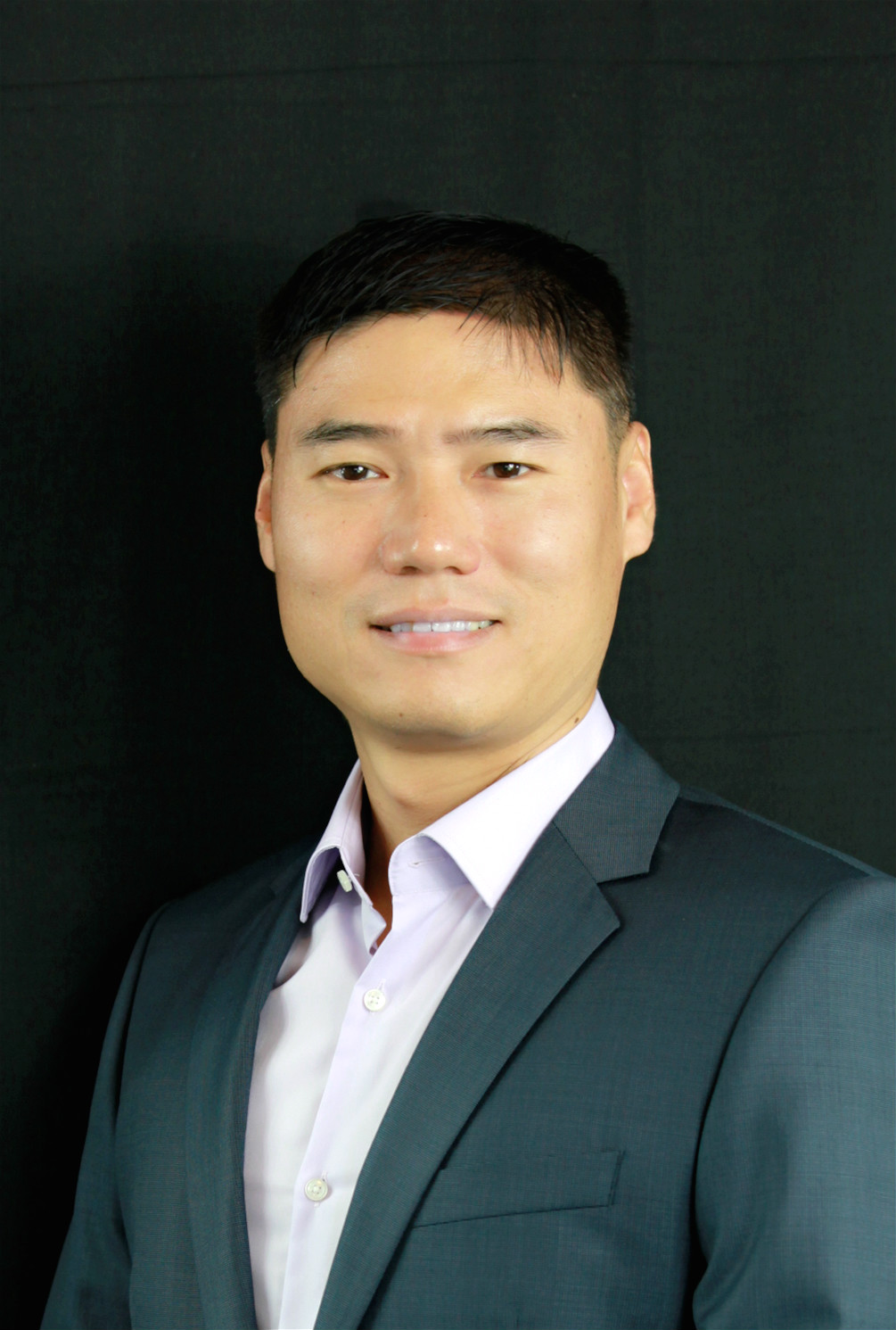}}]{Ying Xu}
 is a postdoctoral researcher at the Department of Electrical and Computer Engineering, University of
Central Florida (UCF), USA. He received the B.Eng, M.Eng and PH.D. degrees from Harbin Institute of Technology, China, in 2003, 2005 and 2009 respectively. From 2009-2017, he has been a Senior Engineer in North China Grid Dispatching and Control Center. His main research interests and experiences include power system analysis, system modeling and control, big-data implementation in power systems, cooperative control and distributed optimization for networked systems.
\end{IEEEbiography}


\begin{IEEEbiography}
[{\includegraphics[width=1in,height=1.25in,clip,keepaspectratio]{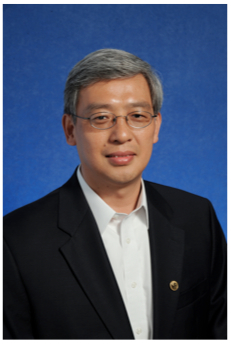}}]{Zhihua Qu} (M’90-SM’93-F’09) received the Ph.D. degree in Electrical Engineering from the Georgia Institute of Technology, Atlanta, in June 1990. Since then, he has been with the University of Central Florida (UCF), Orlando. Currently, he is the SAIC Endowed Professor in College of Engineering and Computer Science, a Pegasus Professor and the Chair of Electrical and Computer Engineering, and the Director of FEEDER Center
(one of DoE-funded national centers on distributed technologies and smart grid). His areas of expertise are nonlinear systems and control, resilient and cooperative control, with applications to energy and power systems.
\end{IEEEbiography}


\begin{IEEEbiography}
[{\includegraphics[width=1in,height=1.25in,clip,keepaspectratio]{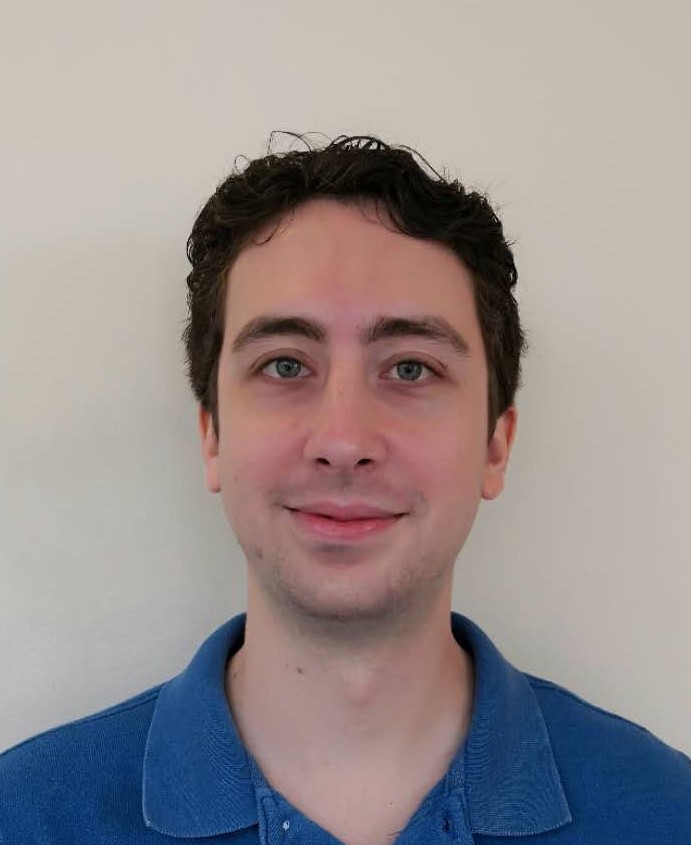}}]{Roland Harvey} is currently pursuing his Ph.D. degree in the Department of Electrical and Computer Engineering, University of Central Florida (UCF), USA. He earned his B.S. from the Department of Physics and Engineering Physics from Tulane University. His research interests include system modeling and control, cooperative control and distributed optimization for networked cyber-physical systems.
\end{IEEEbiography}


\begin{IEEEbiography}
[{\includegraphics[width=1in,height=1.25in,clip,keepaspectratio]{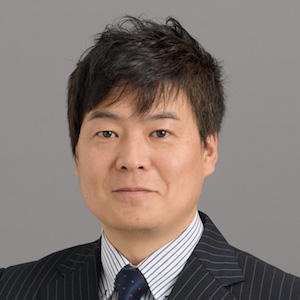}}]{Toru Namerikawa}  (M’94) received the B.E., 	M.E., and Ph.D. degrees in electrical and computer engineering from Kanazawa University, Kanazawa, 	Japan, in 1991, 1993, and 1997, respectively. He was as an Assistant Professor with Kanazawa University, from 1994 to 2002; with the Nagaoka University of Technology, Nigata, Japan, from 2002 to 2005; and again with Kanazawa University, from 2006 to 2009. In 2009, he was with Keio University, Yokohama, Japan, where he is currently a Professor with the Department of System Design Engineering. He held visiting positions with the Swiss Federal Institute of Technology, Zurich,  Switzerland, in 1998; the University of California at Santa Barbara, Santa Barbara, CA, USA, in 2001; the University of Stuttgart, Stuttgart, Germany, in 2008; and Lund University, Lund, Sweden, in 2010. His current research interests include robust control, and distributed and cooperative control and their application to power network systems.
\end{IEEEbiography}

\end{document}